# Direct transition from a disordered to a multiferroic phase on a triangular lattice

M. Kenzelmann[1,2,3], G. Lawes[4,5], A.B. Harris[6], G. Gasparovic[2,3], C. Broholm[2,3], A.P. Ramirez[4,7], G.A. Jorge[8], M. Jaime[8], S. Park[3,9], Q. Huang[3], A.Ya. Shapiro[10], and L.A. Demianets[10]

(1) Laboratory for Solid State Physics, ETH Zurich, CH-8093 Zurich, Switzerland

(2) Department of Physics and Astronomy, Johns Hopkins University, Baltimore, Maryland 21218, USA

(3) NIST Center for Neutron Research, National Institute of Standards & Technology, Gaithersburg, Maryland 20899, USA

(4) Los Alamos National Laboratory, Los Alamos, New Mexico 87545, USA

(5) Wayne State University, Department of Physics & Astronomy, Detroit, Michigan 48202, USA

(6) University of Pennsylvania, Department of Physics & Astronomy, Philadelphia, Pennsylvania 19104, USA

(7) Bell Laboratories, Lucent Technologies, Murray Hill, New Jersey 07974, USA

(8) MPA-NHMFL, MS E536, Los Alamos National Laboratory, Los Alamos, New Mexico 87545, USA

(9) HANARO Center, Korea Atomic Energy Research Institute, Daejeon, Korea

(10) A.V. Shubnikov Institute for Crystallography RAS, 117333 Moscow, Russia

**Competing interactions and geometric frustration provide favourable conditions for exotic states of matter. Such competition often causes multiple phase transitions as a function of temperature and can lead to magnetic structures that break inversion symmetry, thereby inducing ferroelectricity [1-4]. Although this phenomenon is understood phenomenologically [3-4], it is of great interest to have a conceptually simpler system in which ferroelectricity appears coincident with a *single* magnetic phase transition. Here we report the first such direct transition from a paramagnetic and paraelectric phase to an incommensurate multiferroic in the triangular lattice antiferromagnet RbFe(MoO$_4$)$_2$ (RFMO). A magnetic field extinguishes the electric polarization when the symmetry of the magnetic order changes and ferroelectricity is only observed when the magnetic structure has chirality and breaks inversion symmetry. Multiferroic behaviour in RFMO provides a theoretically tractable example of ferroelectricity from competing spin interactions. A Landau expansion of symmetry-allowed terms in the free energy demonstrates that the chiral magnetic order of the triangular lattice antiferromagnet gives rise to a pseudoelectric field, whose temperature dependence agrees with that observed experimentally.**

Magnetic and electric degrees of freedom have different symmetry and their coupling has intrigued scientists continually for over a hundred years [5,6]. Magnetic and electric orders compete on a local scale, making their coexistence relatively rare [7]. While traditional multiferroics have separate magnetic and electric phase transitions [8] a new class of materials with simultaneous magnetic and ferroelectric phase transitions was recently discovered [1-3]. A common feature of the new multiferroics has been spontaneously modulated magnetism with a length scale that is independent from that of the chemical lattice and therefore is called incommensurate. Examples of such materials include the rare-earth (Re) manganites (ReMnO$_3$ and ReMn$_2$O$_5$) [1,2,4] and Ni$_3$V$_2$O$_8$ [3]. The unfamiliar electromagnetic responses of these systems promise fascinating engineering applications if analogous materials or meta-materials that are multiferroic at room temperature can be identified.

Relatively little is known about the microscopic origin of strong coupling between magnetic and ferroelectric order and the role of incommensurability in magneto-electrics. Important insights were provided by a phenomenological theory which was based on a symmetry analysis of the magnetic and ferroelectric order parameters [3,4] and established that the magneto-electric coupling has a trilinear form. A number of microscopic theories have been proposed that respect these symmetry constraints for the known multiferroic materials [9-11]. As we shall see, the present results for RFMO provide a critical test for these theories of magnetoelectricity and reveal that incommensurability is not an essential ingredient.

RFMO is described by space group $P\bar{3}m1$ at room temperature. At $T_0$=180 K the symmetry is lowered to $P\bar{3}$ [12] by a lattice distortion with an amplitude that we denote by $\eta$. Figure 1a shows this structure which, though distorted, still preserves inversion symmetry and therefore is not ferroelectric. For $T < T_0$, RFMO contains perfect $Fe^{3+}$ triangular lattice planes in which spins S=5/2 are coupled through antiferromagnetic superexchange interactions involving two oxygen anions. The magnetism is dominated by interactions within the plane with a relatively small energy scale of about 1 meV [13]. The $Fe^{3+}$ ions feature a sizeable XY-anisotropy which makes spin fluctuations out of the plane unfavourable [13], so that magnetism in RFMO is essentially that of an XY antiferromagnet on a triangular lattice. The planes are coupled via at least 25 times weaker nearest and next-nearest neighbour interactions, inducing magnetic long-range order at $T_N$ =3.8K. Although RFMO contains perfect triangular $Fe^{3+}$ lattices, Fig. 1a shows that out-of-plane ions lead to two types of triangles: "up triangles" with a green oxygen tetrahedron above the plane and "down triangles" with a tetrahedron below the plane. This feature of RFMO implies the absence of a mirror plane perpendicular to the $c$-axis, which is one of the necessary conditions for magneto-electricity.

The zero field magnetic order shown in Fig. 1b is characterized by incommensurate magnetic Bragg peaks that occur at $\mathbf{Q} = \mathbf{G} \pm \mathbf{q}$ in reciprocal space, where $\mathbf{G}$ is a vector of the reciprocal lattice and $\mathbf{q}=(1/3,1/3,q_z)$ is the magnetic ordering wave-vector in reciprocal lattice units. Fig. 2c shows that $q_z \approx 0.458$ for $T < T_N$ in zero field. The temperature dependence of the magnetic Bragg peak intensity, shown in Fig. 2a, is used to identify the onset of magnetic order as a continuous phase transition. At zero field, neighbouring spins in the plane are rotated by $2\pi/3$ with respect to each other in a structure that is commonly referred to as the "120° structure". Spins in adjacent planes are rotated through an angle $2\pi q_z$. There are in fact only two distinct 120° structures, one of which is shown in Fig 1b, and which differ in the spin chirality. An object or a system is said to have chirality if it differs from its mirror image. In RFMO, chirality of the magnetic structures is defined as follows: If, as one traverses clockwise the three vertices of "up triangles" (see Fig. 1b), the spin directions are rotated clockwise by 120° (-120°), then the spin triangle is said to have positive (negative) chirality. The two structures with opposite chirality are represented by two different order parameters $\sigma^{(1)}$ and $\sigma^{(2)}$. These orderings are degenerate in energy, so there will be two types of domains, but within a single domain only one chiral order parameter is nonzero at zero magnetic field.

Due to the small energy scale for magnetic interactions, magnetic fields below 10 T in the basal plane and perpendicular to the $a$-axis have a profound effect on the magnetic structure. At 2.8 K, the low-field incommensurate order collapses in a discontinuous (first order) phase transition for $\mu_0 H = 3.2$ T (see Fig. 2b). For 3.2 T $< \mu_0 H <$ 9.5 T, commensurate order characterized by an ordering vector $\mathbf{q}=(1/3,1/3,1/3)$ is stabilized. For $\mu_0 H \gtrsim 9.5$ T, the magnetic order again becomes incommensurate, and the magnetic periodicity is noticeably field dependent (see Fig. 2d). The phase diagram in Fig. 3 was compiled from the temperature and field dependence of magnetic Bragg peaks, and by tracing specific heat and dielectric anomalies.

Symmetry requires (see Methods) that at nonzero field the magnetic structures in RFMO contain both order parameters $\sigma^{(1)}$ and $\sigma^{(2)}$ of opposite chirality. At ($\mu_0 H$=6 T, $T$=2 K), best agreement with the diffraction data is found for the magnetic structure shown in Fig. 1c where the magnitude of the ordered moments are 3.8(3) $\mu_B$ and 2.8(3) $\mu_B$ along and opposite to the magnetic field, respectively. For the magnetic structure at 10 T, where we have less accurate experimental data, we imposed an equal moment magnitude condition. The best agreement with diffraction data at ($\mu_0 H$=10 T, $T$=100 mK) was found for the magnetic structure shown in Fig. 1d. Fig. 1b-d shows that the incommensurate low-field structure features a chiral order parameter, while both high field commensurate structures contain a point of inversion.

The absence of a point of inversion for the low field magnetic structure prompted us to examine the dielectric properties of the material. Fig. 2a shows the development of ferroelectric polarization along the *c*-axis upon cooling the sample in zero field. Only a single peak was observed in the temperature dependence of the specific heat at zero field [14], demonstrating that the ferroelectric transition coincides with the onset of antiferromagnetic order. Application of a magnetic field reduces the polarization and in the commensurate phase the electric polarization vanishes. Further, we observed small anomalies in the dielectric constant upon entering the commensurate phase as a function temperature and magnetic field. These measurements identify the low-field incommensurate phase as ferroelectric, and the field-induced commensurate phase as paraelectric.

In multiferroics such as TbMnO$_3$ and Ni$_3$V$_2$O$_8$, a ferroelectric order parameter appears when inversion symmetry is broken by the magnetic structure [3,4]. This occurs through two consecutive continuous phase transitions, each characterized by the appearance of an additional magnetic order parameter. A Landau theory [3,4] which couples the two magnetic and one ferroelectric order parameters successfully describes the direction of the ferroelectric polarization, demonstrating that magnetic order induces ferroelectric polarization. Ferroelectricity in these systems may also be understood by the spin-current interaction which is due to superexchange and the spin-orbit interactions [9], and which gives the spontaneous electric polarization that arises from canted spins on neighbouring sites $i$ and $j$ as $\mathbf{P}_{ij}$ = [$\mathbf{S}(\mathbf{r}_i)$ x $\mathbf{S}(\mathbf{r}_j)$] x [$\mathbf{r}_i$ - $\mathbf{r}_j$]. For RFMO $\mathbf{P}_{ij}$ lies in the basal plane and, in view of the three-fold rotation axis, the corresponding macroscopic polarization P=$\Sigma_{<ij>}$ $\mathbf{P}_{ij}$ vanishes. Since RFMO does exhibit ferroelectricity, the spin-current interaction alone [9], while possibly valid for some systems, does not provide a general explanation for magneto-electricity. In contrast, we shall show that the symmetry based penomenological theory [3,4,15] is fully consistent with the new experimental results in RFMO.

The direct multiferroic transition in RFMO and the fact that there is just one magnetic atom per unit cell greatly simplifies the analysis and reduces the number of parameters in a phenomenological symmetry based theory. Here we will explicitly consider only the case of zero magnetic field, although the general conclusions are expected to remain valid over the entire low-field phase until a phase boundary is crossed. The most general form that the antiferromagnetic spin distribution can take is

$$S_x(\mathbf{r}) = [\sigma^{(1)}(q_z) + \sigma^{(2)}(q_z)]e^{i\mathbf{q}\cdot\mathbf{r}} + \text{complex conjugate}$$

$$S_y(\mathbf{r}) = -i[\sigma^{(1)}(q_z) - \sigma^{(2)}(q_z)]e^{i\mathbf{q}\cdot\mathbf{r}} + \text{complex conjugate} , \qquad (1)$$

where $\sigma^{(n)}(q_z)$ (which are complex-valued) are the previously mentioned magnetic order parameters. **q** has an in-plane component (1/3,1/3,0) which generates a 120° spin structure in each triangular lattice plane, and an out-of-plane component, $q_z$, which describes the interplanar spin rotation in the ordered structure. In terms of these order parameters the free energy at quadratic order in zero magnetic field is

$$F = \frac{1}{2}\sum_{n,\pm} r(\pm q_z)|\sigma^{(n)}(\pm q_z)|^2 + O(\sigma^4) \qquad (2)$$

where $r(q_z)$ is an even function of $q_z$ in the $P\bar{3}m1$ space group above $T_0$=180 K, but develops a contribution proportional to $\eta\, q_z$ in the $P\bar{3}$ space group below $T_0$. Minimization of $F$ shows that $1/2-q_z$ is proportional to the lattice distortion $\eta$ away from the $P\bar{3}m1$ space group. The transition at $T_0$=180 K does not break inversion symmetry, but it does determine the incommensurate wave-vector $q_z$. As Eq. (2) indicates, the lattice distortion $\eta$ actually selects the helicity (the sign of $q_z$), although it does not lift the degeneracy between magnetic structures with opposite chirality $\sigma^{(1)}$ and $\sigma^{(2)}$. At zero field the effect of the terms in the free energy of order $\sigma^4$ (which we will not discuss in detail) is to prevent $\sigma^{(1)}$ and $\sigma^{(2)}$ from simultaneously being nonzero. We now consider the lowest order terms allowed by symmetry which couple the $\sigma$ order parameters describing magnetism to the spontaneous polarization **P** which describes ferroelectricity. The lowest order interaction that induces a spontaneous polarization and conserves wave-vector must be of the form $V = \Sigma_{nm\gamma} c^{\gamma}_{nm} \sigma^{(n)}(q_z)\sigma^{(m)}(q_z)^* P_\gamma$, where $\gamma$ labels the cartesian component of the electric polarization. The fact that $V$ must be invariant under the spatial inversion operator $\mathcal{I}$ implies that $c^{\gamma}_{nm}$ is only nonzero for $n=m$ (see Methods). Furthermore, since $|\sigma^{(n)}(q_z)|^2$ is invariant under the three-fold rotation about the $c$-axis $\mathcal{R}_3$, one must have $\mathcal{R}_3 P_\gamma = P_\gamma$. Thus the magneto-electric free energy in RFMO assumes the form

$$F = \frac{1}{2}\chi^{-1}P^2 + K\left[\left|\sigma^{(1)}(q_z)\right|^2 - \left|\sigma^{(2)}(q_z)\right|^2\right]P_c \qquad (3)$$

where $\chi$ is the electric susceptibility, which we have measured to be $\chi \approx 4(1)\,\varepsilon_0 \approx 4(1)$ $\cdot 10^{-11}$ C$^2$/Jm away from the phase boundaries and the value of the coefficient $K$ is not fixed by symmetry. In contrast to the situation in some of the previously studied oxides [1-4], the present results show that ferroelectricity induced by magnetism does not require the simultaneous presence of two separate magnetic order parameters that are critical at distinct continuous phase transitions.

An advantage of the symmetry based phenomenological theory is its close connection to microscopic models, such as the magneto-elastic theory developed for Ni$_3$V$_2$O$_8$ [11]. At low temperature, when $\sigma$ is of order unity, Fig. 2a shows an electric polarization $P \approx 5 \cdot 10^{-6}$ C/m$^2$ from which we deduce that $K \approx 2 \cdot 10^5$ V/m. This value of $P$ corresponds to a ferroelectric atomic displacement (probably of the oxygen ions) of order $10^{-5}$Å. The trilinear interaction suggests that ferroelectricity could be induced by a microscopic interaction of the form $V \sim (\partial J_{\alpha\beta}/\partial u_{ij})\, u_{ij}\, S_\alpha(i)\, S_\beta(j)$, where $J_{\alpha\beta}$ is an exchange tensor. The value of $K$ given above corresponds to $\partial J_{\alpha\beta}/\partial u$ of order $10^{-4}$ eV/Å for $\alpha \neq \beta$.

When Eq. (3) is minimized with respect to $P_c$, the electric polarization along the $c$-axis, one finds its equilibrium value, denoted $<P_c>$, to be proportional to $|\sigma^{(1)}|^2 - |\sigma^{(2)}|^2$. So $P_c$ is nonzero as long as the net chirality on "up triangles" $|\sigma^{(1)}|^2 - |\sigma^{(2)}|^2$ is nonzero. At zero field, only one of the two magnetic order parameters $\sigma^{(1)}$ and $\sigma^{(2)}$ is non-zero, so the temperature dependence of the magnetic Bragg peak, which is proportional to $|\sigma^{(i)}|^2$, is predicted by this theory to exactly match that of the ferroelectric polarization. The excellent agreement of the data reported in Fig. 2a with this prediction provides strong support for the validity of the theory. The fact that $P_c$ couples to a difference in order parameters suggests that applying an electric field to an un-twinned sample at zero magnetic field, would favour a magnetic structure with definite chirality on "up triangles" described by $\sigma^{(1)}$ or $\sigma^{(2)}$. Furthermore because there is a definite helicity associated with the incommensurate modulation, an untwinned sample in an electric field should also be helical.

Both commensurate magnetic structures found in applied field of 6 T and 10 T have $|\sigma^{(1)}|^2=|\sigma^{(2)}|^2$ within error bars, so there can be no magneto-electric coupling in those phases. This is consistent with the observed absence of ferroelectricity there. Eq. (3) further implies that the high-field incommensurate phase is ferroelectrically polarized along the *c*-axis if $|\sigma^{(1)}|^2 \neq |\sigma^{(2)}|^2$ and paraelectric otherwise. While these results seem to indicate otherwise, incommensurability is not a requirement for magneto-electricity in RFMO. Indeed we claim that even if the lattice distortion were zero, the resulting commensurate magnetic structure would still be ferroelectric. This can be appreciated through Landau theory or through the following intuitive argument. To induce a ferroelectric polarization a magnetic structure must break inversion symmetry and define a unique direction. Fig. 1b shows that the 120º structure clearly breaks inversion symmetry. To see that it defines a direction we provide the following prescription to specify a unique direction in the magnetically ordered crystal. Identify triangles with positive chirality. Even in the absence of the distortion $\eta$, the direction from such a triangle to the nearest oxygen tetrahedron (green triangles in Fig. 1a) defines a unique direction. So while the absence of a mirror plane perpendicular to the *c*-axis is an essential ingredient of magnetoelectricity, incommensurability surprisingly is not.

In summary, the novel multiferroic RFMO with a simultaneous onset of magnetic and ferroelectric order offers the simplest case thus far wherein a phenomenological trilinear coupling theory accurately describes the symmetry properties of the magnetic and ferroelectric order parameters. Our approach allows quantitative comparisons between theory and experiment and leads to the astonishingly general prediction that trigonal stacked triangular lattice antiferromagnets with a 120° spin structure are multiferroic!

**Methods**

Single-crystal samples of RFMO were grown by a flux method described in Ref. 16. Neutron measurements were carried out using the BT2 and SPINS instruments at NIST and a superconducting magnet with a dilution refrigerator. The temperature and field dependence of the magnetic scattering was measured using BT2 with an incident energy of $E_i$=14.7 meV and SPINS with an incident energy of $E_i$=5 meV. The magnetic structures were determined by measuring a number of magnetic Bragg peaks and comparing their relative intensities to those calculated for symmetry allowed structures. The magnetic structure at zero field was determined using 82 magnetic Bragg peaks measured with BT2 and $E_i$=35meV. The magnetic structure at $\mu_0 H$=6T and $T$=2K was determined from 62 magnetic Bragg peaks measured using BT2 with $E_i$=30meV. The magnetic structure at $\mu_0 H$=10T was determined by measuring 15-30 magnetic Bragg peaks using SPINS with $E_i$=5meV. The agreement between the model and the data is given by $\chi^2$=10.6 and the R-factor $R$=0.11 at zero field, $\chi^2$=5.2 and $R$=0.14 at $\mu_0 H$=6T and $\chi^2$=22.6 and $R$=0.21 at $\mu_0 H$=10T. Here $\chi^2$ is the mean squared deviation between model and data in units of the variance and $R=(1/N)\,\Sigma_i|I^o{}_i - I^c{}_i| / I^o{}_i$ ($I^o{}_i$ and $I^c{}_i$ are the observed and calculated intensities, respectively). The magnetic structure at 6T is described by $\sigma^{(1)} =$ (1.09(4) + i 1.85) and $\sigma^{(2)} =$ (1.09(4) + i 1.94(4)), and the magnetic structure at 10T where we imposed an equal spin length condition by $\sigma^{(1)} =$ (2.16(3) + i 0.197) and $\sigma^{(2)} =$ (-1.251 - i 1.724) (error bars are only provided for free parameters). To probe ferroelectric order, we sputtered two gold electrodes on opposite faces of the crystal plates (the crystals grow in platelets perpendicular to the *c*-axis) and measured both the pyroelectric current at fixed *H*, and the magnetoelectric current at fixed *T*, using a Keithley electrometer. We then integrated the current to find the spontaneous polarization *P* in the ferroelectric state. We aligned the ferroelectric domains by applying a (~ 2 kV/cm) polarizing electric field $E_0$ as the sample was cooled through the transition temperature. Specific heat measurements where performed on a 2 mg single crystal sample with a SiN thin-membrane microcalorimeter. The thermal relaxation technique was used [17], with a NbSi composite thermometer and a Pt heater deposited on the back of the SiN membrane and calibrated in-situ. The magnetic field produced by a 11.5T superconducting magnet was applied in the crystallographic *ab*-plane. Group theory was used to determine all allowed magnetic structures. The group $\mathcal{G}$ of symmetry elements that leave **q** invariant consists of three elements and is generated by powers of $\mathcal{R}_3$, the three-fold rotation. There are three possible irreducible representations: One, which the experiment clearly excludes, has the spin moments aligned along the *c*-axis. The other two allow only in-plane order. Analysis of the quadratic terms in the Landau expansion in powers of the Fourier components of the spin distribution indicates that if the phase is reached from the paramagnetic phase via a continuous transition, then the vector Fourier coefficients must be eigenvectors of $\mathcal{R}_3$ with eigenvalue either $\lambda_1$ or $\lambda_2$, where $\lambda_2=\lambda_1^2=\exp(4\pi i/3)$. The Fourier representation of the spin distribution involves both the wavevectors **q** and −**q**, and the most general form the spin distribution can take is given by Eq. (1). Using $\mathcal{I}\mathbf{r} = -\mathbf{r}$ one can further show that $\mathcal{I}\sigma^{(1)}(q_z)= \sigma^{(2)}(q_z)^*$, where $\mathcal{I}$ is the spatial inversion operator. In addition we obviously have $\mathcal{I} P_\gamma = - P_\gamma$. For the magneto-electric interaction $V= \Sigma_{nm\gamma} c^\gamma_{nm}\sigma^{(n)}(q_z)\sigma^{(m)}(q_z)^* P_\gamma$, invariance under $\mathcal{I}$ implies that $c^\gamma_{nm}$ is only nonzero for n=m and $c^\gamma_{11} = - c^\gamma_{22}$.

1. T. Kimura et al, Nature (London) **426**, 55 (2003).

2. N. Hur et al, Nature (London) **429**, 392 (2004).

3. G. Lawes et al, Phys. Rev. Lett. **95**, 087205 (2005).

4. M. Kenzelmann et al, Phys. Rev. Lett. **95**, 087206 (2005).

5. P. Curie, J. Physique **3,** 393, (1894).

6. M. Fiebig, J. Phys. D: Appl. Phys. **38,** R123-R152 (2005).

7. N.A. Hill et al, J. Magn. Magn. Mat. **242**, 976 (2002).

8. G.A. Smolenskii and I.E. Chupis, Sov. Phys.—Usp. **25,** 475 (1982).

9. H. Katsura et al, Phys. Rev. Lett. **95**, 057205 (2005).

10. I.A. Sergienko and E. Dagotto, Phys. Rev. B **73**, 094434 (2006).

11. A. B. Harris et al, Phys. Rev. B **73**, 184433 (2006).

12. G. Gasparovic, Ph.D. Thesis, Johns Hopkins University.

13. L.E. Svistov et al, Phys. Rev. B **67**, 094434 (2003).

14. G.A. Jorge et al, Physica B **354**, 297 (2004).

15. M. Kenzelmann et al, Phys. Rev. B **74**, 014429 (2006).

16. R.F. Klevtsova and P.V. Klevtsov, Kristallografiya **15**, 209 (1970).

17. R. Bachmann, et al, Rev. Sci. Instrum. **43**, 205 (1972).

**Acknowledgements** Work at ETH was supported by the Swiss National Science Foundation under Contract No. PP002-102831. Work at the National High Magnetic Field Laboratory was supported by the U.S. National Science Foundation through Cooperative Grant No. DMR901624, the State of Florida, and the U.S. Department of Energy. Work at the Shubnikov Institute was supported by the US Civilian Research and Development Foundation Grant No. RP1-2097. Work at Johns Hopkins University was supported by the National Science Foundation through Grant No. DMR-0306940. The work at SPINS was supported by NSF through DMR-9986442.

**Competing interest statement** The authors declare that they have no competing financial interests.

**Correspondence** and requests for materials should be addressed to M. Kenzelmann (kenzelmann@phys.ethz.ch).

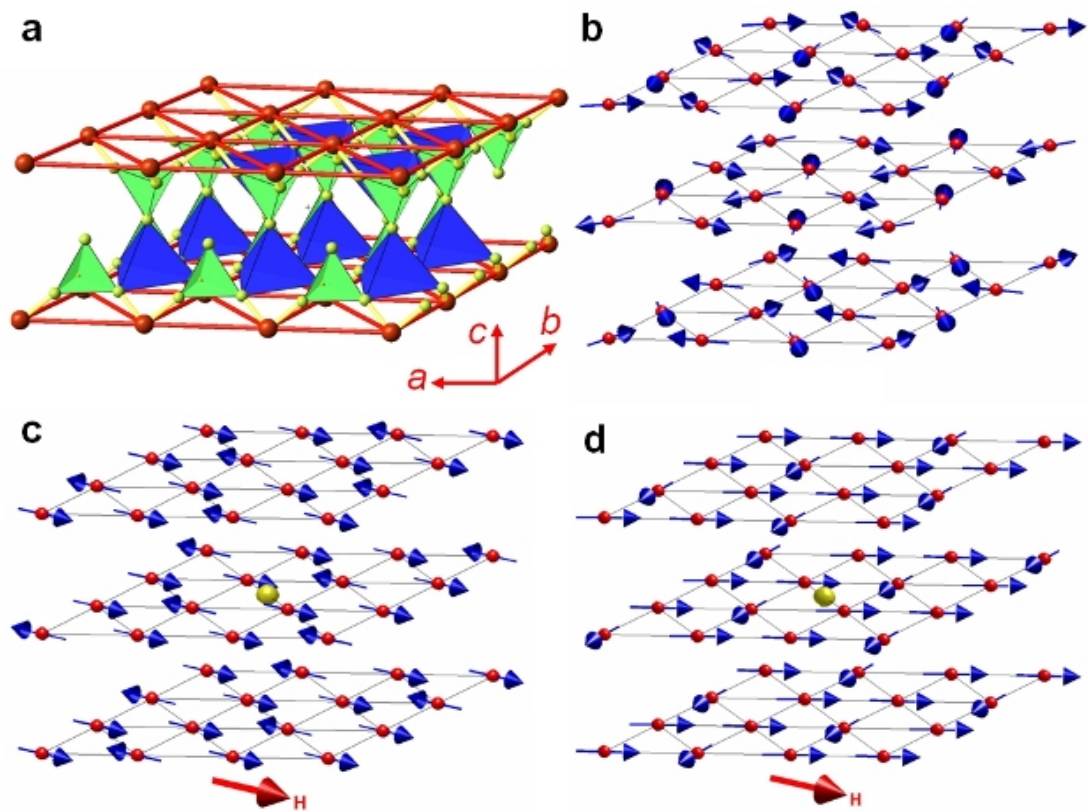

**Figure 1 a:** Low-temperature chemical structure of RFMO, belonging to the trigonal space group $P\bar{3}$. Shown are the $Fe^{3+}$ in red and $O^{-2}$ ions in yellow. The $Fe^{3+}$ are expected to interact through superexchange involving two $O^{-2}$ for intra-plane interactions, and three $O^{-2}$ for inter-plane interactions. There is one nearest-neighbour and two distinct next-nearest neighbour interactions between planes. Magnetic structures **b**: at $\mu_0 H=0$T and $T=2$K with an ordered moment at each site $M=3.9(5)$ $\mu_B$, **c**: at $\mu_0 H=6$T and $T=2$K is a collinear spin arrangement where two spins point with **M**=3.8(3) $\mu_B$ along the field direction and one spin with $M$=2.8(3) $\mu_B$ opposite to it, and **d**: at $\mu_0H$=10T and $T$=0.1K with $M$ = 3.7(5) $\mu_B$ is a non-collinear spin arrangement where one third of the spins are perpendicular to the field and two thirds of the spins are parallel to each other and form an angle of thirty degrees with the field direction. The structures shown in **c**-**d** have an inversion centre which is indicated by the yellow point, and the zero-field magnetic structure in b features a chiral order parameter.

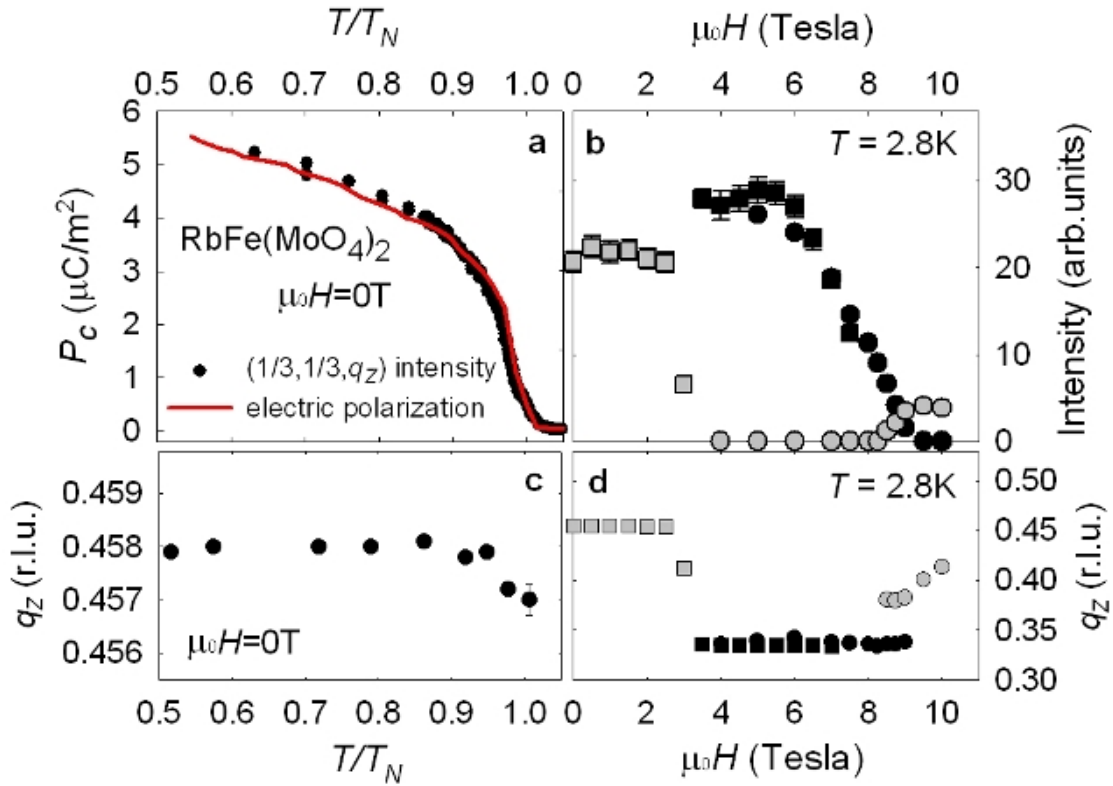

**Figure 2 a:** Zero-field temperature dependence of the magnetic Bragg intensity observed at **Q**=(1/3,1/3,$q_z$), compared to the temperature dependence of the ferroelectric polarization $P_c$ along the $c$-axis. **b**: Field dependence of the intensity at $T$=2.8K for the commensurate (black) and the incommensurate (grey) reflection. The commensurate and high-field incommensurate order coexist at this temperature for a narrow field region between $\mu_0H$ = 8 and 9 T. **c**:

Temperature dependence of the incommensurate magnetic wave-vector $q_z$ at zero field. **d**: Field dependence of $q_z$ at *T*=2.8K, shown for the commensurate (black) and incommensurate (grey) magnetic Bragg reflections. Squares and circles in b,d distinguish two independent measurements, for which intensities were put on the same scale by matching data measured at the same field.

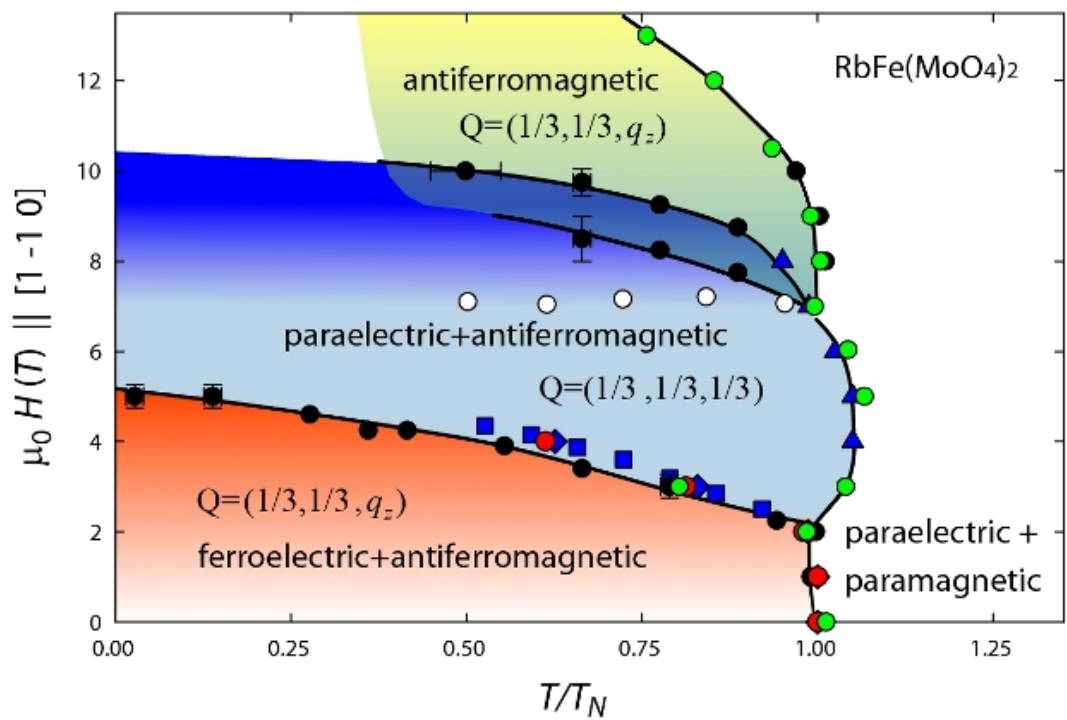

**Figure 3** *H-T* phase diagram of RFMO, for magnetic field applied along the [1 -1 0] direction, obtained through neutron diffraction (solid black circles), specific heat (solid green circles), dielectric (solid blue squares represent peaks as a function of field, solid blue diamonds and triangles indicate anomalies as a function of temperature), and pyroelectric measurements (solid red circles). The ordering temperature $T_N \approx 3.8$ K. Open circles represent the end of a magnetization plateau as a function of field [7]). The color gradient indicates the estimated increasing magnetization as a function of magnetic field. The boundary of the high-field incommensurate phase has not been determined for

$T < 2K \approx 0.5 T_N$. Subtle phase modifications which may occur in small regions adjacent to the **Q**=(1/3,1/3,1/3) phase [7] are not shown. The commensurate and the high-field incommensurate magnetic order coexist in a narrow field region for $\mu_0 H \approx 9$ T. The two solid lines indicate where each type of order was observed in diffraction data (see Figs. 2b and 2d).